\documentstyle[preprint,aps,epsf]{revtex}

\newcommand{\beq}{\begin{equation}}
\newcommand{\eeq}{\end{equation}}
\def\bea{\begin{eqnarray}}  \def\eea{\end{eqnarray}}

\begin{document}

\begin{center}
\vspace*{1 truecm}
{\Large \bf 
Multiplicity fluctuations in the string clustering approach}\\[8mm]
{\bf  L. Cunqueiro, E. G. Ferreiro, F. del Moral and C. Pajares}\par
{\it Departamento de
F\'{\i}sica de  Part\'{\i}culas and Instituto Galego de F\'{\i}sica de Altas
Energ\'{\i}as,
Universidade de Santiago de
Compostela,\\ 15782--Santiago de Compostela, Spain}

\vskip 1.0truecm
{\Large {\bf Abstract}}
\end{center}
\begin{quotation}
We present our results on multiplicity fluctuations in the framework of the
string clustering approach. 
We compare our results --with and without clustering formation-- with CERN SPS NA49 data.
We find a non-monotonic behaviour of these fluctuations
as a function of the collision centrality, which has the same origin as the observed
fluctuations of transverse 
momentum: the correlations between the produced particles due to the cluster formation.
\end{quotation}
\vskip 1.0truecm


\vspace{5cm}

\newpage

Non-statistical event-by-event fluctuations have been proposed as a possible signature 
for QCD phase
transition. In a thermodynamical picture of the strongly interacting system formed in 
heavy-ion
collisions, the fluctuations of the mean transverse momentum or mean multiplicity 
are related to the fundamental properties of the system, such the specific heat, so may
reveal information about the QCD phase boundary.

Event-by-event fluctuations of transverse momentum have been measured both at CERN SPS 
\cite{Appelshauser:1999ft,Anticic:2003fd,Blume:2003,Gazdzicki2004,RybczynskiNA49,Adamova:2003pz}
and BNL RHIC 
\cite{Adcox:2002pa,Adler:2003xq,Adams:2003uw,Pruneau:2004nc,Mitchell:2004xz}. 
The data show a non-trivial behaviour as a function of the centrality of the
collision. 
Concretely, the non-statistical normalized fluctuations grow as the centrality increases, with a maximum
at mid centralities, followed by a decrease at larger centralities.
Different mechanisms 
\cite{Gazdzicki:1997fg,Liu:1998xf,Bleicher:1998wu,Mrowczynski:1998vt,Stephanov:1999zu,Capella:1999uc,Baym:1999up,Dumitru:2000in,Korus:2001fv,Korus:2001au,Stephanov:2001zj,Voloshin:2002ku,Gavin:2003cb,Rybczynski:2003jk,Ferreiro:2003dw,DiasdeDeus:2003sn,Liu:2003jf,Stephanov:2004wx}
have been proposed in order to explain those data, including 
complete or partial equilibration \cite{Bleicher:1998wu,Mrowczynski:1998vt,Stephanov:2001zj,Gavin:2003cb}, 
critical phenomena \cite{Stephanov:1999zu,Stephanov:2004wx}, 
production of jets \cite{Liu:1998xf,Liu:2003jf}
and also string clustering or string percolation \cite{Ferreiro:2003dw,DiasdeDeus:2003sn}.

In particular, we have proposed \cite{Ferreiro:2003dw} an explanation for those fluctuations based on the creation of
string clusters. 
In our approach, we find an increase of the mean $p_T$ fluctuations at mid centralities, followed by a
decrease at large centrality. Moreover, we obtain 
a similar behaviour at SPS and RHIC energies.
In the framework of string clustering such a behaviour is naturally explained. As the centrality increases,
the strings overlap in the transverse plane forming clusters. These clusters decay into particles with mean
transverse momentum and multiplicities that depend on the number of strings that conform each cluster.
The event-by-event 
fluctuations on mean $p_T$ and mean multiplicity correspond then to fluctuations of the 
transverse momentum and
multiplicity of those clusters and behave essentially as the number of clusters conformed by a different
number of strings.   
If the number of different clusters --different in this context means that the clusters are made of
different numbers of strings-- grows, that will lead to an increase of fluctuations. And in fact this number
grows with centrality up to a maximum. For higher centralities, the number of different clusters decreases. 

On the other hand, in a jet production scenario
\cite{Adler:2003xq}, the mean $p_T$ fluctuations are attributed to jet production in peripheral events, combined with
jet suppression at larger centralities. A possible way to discriminate between the two approaches could be the study of
fluctuations at SPS energies, where jet production cannot play a fundamental role.

Recently, the NA49 Collaboration have presented their data on multiplicity fluctuations as a
function of centrality \cite{Gazdzicki:2004ef,Rybczy2004} at SPS energies.
In order to develop the experimental analysis, the variance of the multiplicity distribution
$Var(N)=<N^2>-<N>^2$ scaled to the mean value of the multiplicity $<N>$ has been used.
A non-monotonic centrality --system size-- dependence of the scaled variance was found.
In fact, its behaviour is similar to the one obtained for the $\Phi (p_T)$-measure 
\cite{Gazdzicki:ri} used by the NA49 Collaboration to quantify the $p_T$-fluctuations, suggesting 
they are related to each other \cite{Mrowczynski2004,Rybczy2004:9}.
The $\Phi$-measure is independent of the distribution of number of particle sources
if the sources are identical and independent from each other. This implies that $\Phi$ is
independent of the impact parameter if the nucleus-nucleus collision is
a simple superposition of nucleon-nucleon interactions.


Our aim in this note is to 
calculate the event-by-event multiplicity fluctuations 
applying the same mechanism --clustering of colour strings-- that 
we have used previously \cite{Ferreiro:2003dw} for the study of the $p_T$-fluctuations.
Let us remember the main features of our model. In each nuclear collision, colour strings are
streached between partons from the projectile and the target, which decay into new strings by 
sea $q-\bar{q}$ production and finally hadronize to produce the observed particles. For the decay of
the strings we apply the Schwinger mechanism of fragmentation \cite{Schwinger}, where the decay is
controlled by the string tension that depends on the colour charge and colour field of the string.
The strings have longitudinal and transverse dimensions, and the density of created strings in the
first step of the collision depends on the energy and the 
centrality of the collision. Roughly speaking,
one can consider the number of strings $N_s$ in the central rapidity region 
as proportional to the number of collisions, 
$N^{4/3}_A$, while in the forward region it becomes proportional to the number of participants, $N_A$.
We define the density of strings in the transverse space as $\eta=\frac{N_s S_1}{S_A}$,
where $N_s$ is the total number of strings created in the collision, each one
of an area $S_1=\pi r_0^2$ ($r_0 \simeq 0.2 \div 0.3$ fm), 
and $S_A$ corresponds to the nuclear overlap area,
$S_A=\pi R^2_A$ for central collisions. With the increase of energy and/or atomic number of the
colliding nuclei, this density grows, so the strings begin to overlap forming clusters
\cite{Percolation}.

We assume that a cluster of $n$ strings that occupies an area $S_n$
behaves as a single colour source with a higher colour field, generated
by a higher colour charge
$Q_n$. This charge
corresponds to the vectorial sum of the colour charges of each individual
string ${\bf Q}_1$. The resulting colour field covers the area $S_n$ of
the cluster. As $Q_n^2=( \sum_1^n {\bf Q}_1)^2$,
and the individual
string colours may be arbitrarily oriented, 
the average ${\bf Q}_{1i}{\bf Q}_{1j}$ is zero, so
$Q_n^2=n  Q_1^2$ if the strings fully overlap. Since the strings may
overlap only partially we introduce a dependence on the area of the cluster. 
We obtain $Q_{n}= \sqrt{ n S_n \over S_1} Q_{1}$ \cite{19}.
Now we apply the Schwinger mechanism for the fragmentation of the cluster, and one obtains 
a relation between the mean multiplicity $<\mu>_n$ and the average transverse momentum $<p_T>_n$
of the particles produced by a cluster
of $n$ strings that covers an area $S_n$:
\beq
<\mu>_n=\sqrt{\frac{n S_n}{S_1}} <\mu>_1\ \ \ {\rm and}\ <p_T>_n=\Big ( \frac{n S_1}{S_n}
\Big )^{1/4} <p_T>_1 \, ,
\label{ec1}
\eeq
where $<\mu>_1$ and $<{p_T}>_1$ correspond to
the mean multiplicity and the mean transverse momentum of the particles
produced by one individual string.

In order to obtain the mean $p_T$ and the mean multiplicity of the collision at a given centrality, one
needs to sum over all formed clusters and to average over all events:
\beq
<\mu>=\frac{\sum_{i=1}^{N_{events}} \sum_j <\mu>_{{n_j}}}{N_{events}}\, , \, \, 
<p_T>=\frac{\sum_{i=1}^{N_{events}} \sum_j
<\mu>_{{n_j}} <p_T>_{{n_j}}}
{\sum_{i=1}^{N_{events}} \sum_j <\mu>_{{n_j}}} \ .
\label{ec2}
\eeq
The sum over $j$ goes over all individual clusters $j$, each one formed by ${{n_j}}$
strings and occupying an area $S_{{n_j}}$.
The quantities ${{n_j}}$ and $S_{{n_j}}$
are obtained for each event, using a Monte Carlo code \cite{21,22},
based on the quark gluon string model. Each string is generated at an
identified impact parameter in the transverse space. Knowing the transverse area
of each string, we identify all the clusters formed in each event, the
number of strings ${{n_j}}$ that conforms each cluster $j$, and the area
occupied by each cluster $S_{{n_j}}$. Note that for two different clusters, $j$ and $k$, formed by the
same number of strings ${{n_j}}=n_k$, the areas $S_{{n_j}}$ and $S_{n_k}$ can vary.
Because of this we do the sum over all individual clusters.
So we use a Monte Carlo for the cluster formation, in order to compute the
number of strings that come into each cluster and the area of the cluster. On the other hand,
we do not use a Monte
Carlo code for the decay of the cluster, since we apply analytical expressions (eqs. (\ref{ec1})) for the
transverse momentum $<p_T>_{{n_j}}$ and the multiplicity $<\mu>_{{n_j}}$ of each individual cluster.

In order to obtain the scaled variance we calculate $<\mu^2>$:
\beq
<\mu^2>=
\frac{1}{N_{events}} \Big [
\sum_{i=1}^{N_{events}} \Big (\sum_j \sqrt{\frac{n_j S_{n_j}}{S_1}} \Big )^2 <\mu>_1^2 +
\sum_{i=1}^{N_{events}} \sum_j \sqrt{\frac{n_j S_{n_j}}{S_1}} <\mu>_1 \Big ] \ ,
\label{ec3}
\eeq
where we have supposed that the
 multiplicity of each cluster follows a Poissonian of mean
value $<\mu>_{{n_j}}$, and we have applied the property for a Poissonian:\\ 
$<\mu^2>_{{n_j}}=<\mu>^2_{{n_j}}+<\mu>_{{n_j}}$.

Finally, our formula for the scaled variance obeys:
\beq
\frac{Var(\mu)}{<\mu>}= 1+ <\mu>_1
\frac{\Big < \Big (\sum_j \sqrt{\frac{n_j S_{n_j}}{S_1}} \Big )^2 \Big > - \Big < \sum_j \sqrt{\frac{n_j
S_{n_j}}{S_1}} \Big >^2}
{ \Big < \sum_j \sqrt{\frac{n_j
S_{n_j}}{S_1}} \Big >} \ ,
\label{ec4}
\eeq
where the mean value in the r.h.s. corresponds to an average over all events.

The behaviour of this quantity is as follows:
in the limit of low density --isolated strings that do not interact--,
\beq
\frac{Var(\mu)}{<\mu>}= 1+<\mu>_1
\frac{<N_s^2>-<N_s>^2}{<N_s>}
\label{ec5}
\eeq
where $N_s$ corresponds to the number of strings. Considering 
that, for a fixed number of participants, 
the number of strings behaves as a Poissonian distribution we obtain
\beq
\frac{<N_s^2>-<N_s>^2}{<N_s>} \simeq 1 \, ,
\label{ec6}
\eeq
so
\beq
\frac{Var(\mu)}{<\mu>}= 1+<\mu>_1 \, .
\label{ec7}
\eeq
 In the large density regime --all the strings fuse into a single cluster that occupies the whole
interaction area-- we have:
\beq
\frac{Var(\mu)}{<\mu>}= 1+<\mu>_1
\frac{\Big < \Big (\sqrt{\frac{N_s S_A}{S_1}} \Big )^2 \Big > -\Big < \sqrt{\frac{N_s S_A}{S_1}}\Big >^2}
{\Big < \sqrt{\frac{N_s S_A}{S_1}}\Big >} \, ,
\label{ec8}
\eeq
where $S_A$ is the nuclear overlap area. 
The second element of the r.h.s. of this equation tends to
zero, and the scaled variance becomes equal to one.

\begin{figure}
\centering\leavevmode
\epsfxsize=6.5in\epsfysize=6.5in\epsffile{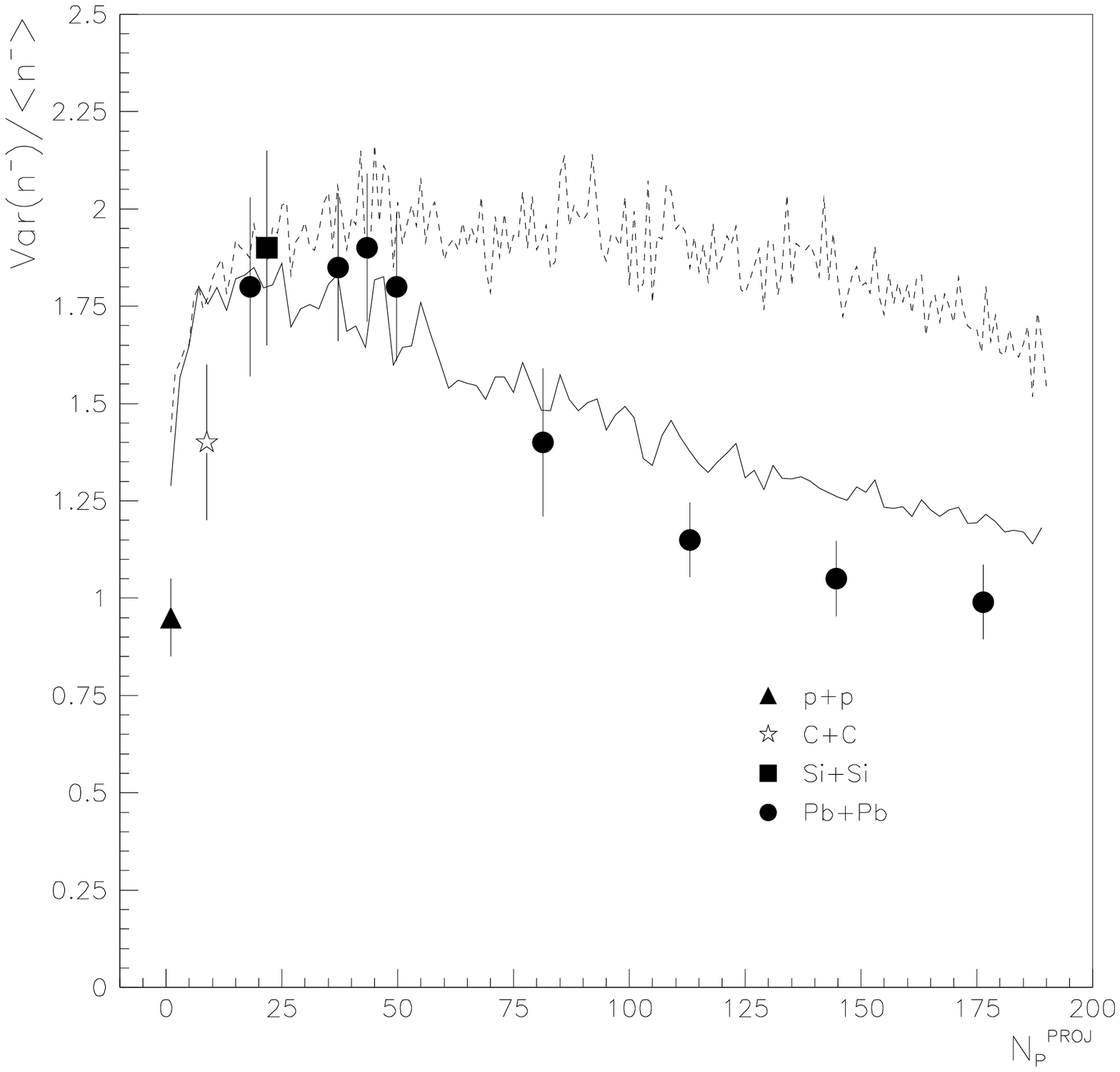}
\vskip 0.5cm
\caption{
Our results for the scaled variance of negatively charged particles in Pb+Pb collisions at 
$P_{lab}=$158 AGeV/c 
compared to NA49
experimental data. The dashed line corresponds to our result when clustering formation
is not included, the continuous line takes into account clustering.}
\label{fig1}
\end{figure}

Our results for the scaled variance for negative particles $\frac{Var(n^-)}{<n^->}$
compared to experimental data 
\cite{Gazdzicki:2004ef,Rybczy2004} are presented in 
Fig. \ref{fig1}. Note that in order to obtain these results we need to fix the value of the parameter
$<\mu>_1$. It is defined as $<\mu>_1=<\mu>_{0} \Delta y$, where $<\mu>_{0}$ 
is the number of particles produced
by one individual string and $\Delta y$ corresponds to the rapidity interval considered. We do not
introduce any dependence of $<\mu>_0$ with the energy or the centrality
of the collision.
The value of $<\mu>_{0}$ has been previously fixed from a comparison of the model to SPS and RHIC data
\cite{19,20}
on multiplicities. In the first Ref. of \cite{19}, the total
multiplicity per unit
rapidity
produced by one string has been taken as $<\mu>_{0\, {tot}} \simeq 1$.
If we assume that
$1/3$ of the created particles are negative, that would lead to a negative
particle multiplicity per unit rapidity for each individual string of
$<\mu>_{0\, {neg}}=0.33$.
The rapidity interval considered, in order to compare with NA49 experimental data, is $4.0 < y < 5.5$.
The data are obtained in a restricted $p_T$ range, $0.005 < p_T < 1.5$ GeV/c, while our results take into
account all possible transverse momenta. Nevertheless, the experimental acceptance covers the small $p_T$
region, which gives the largest contribution at SPS energies. Because of this, we obtain a good agreement for
the centrality dependence of $<p_T>$ (see Table 1 of Ref. \cite{Ferreiro:2003dw} for more details.).

\begin{figure}
\centering\leavevmode
\epsfxsize=6in\epsfysize=6in\epsffile{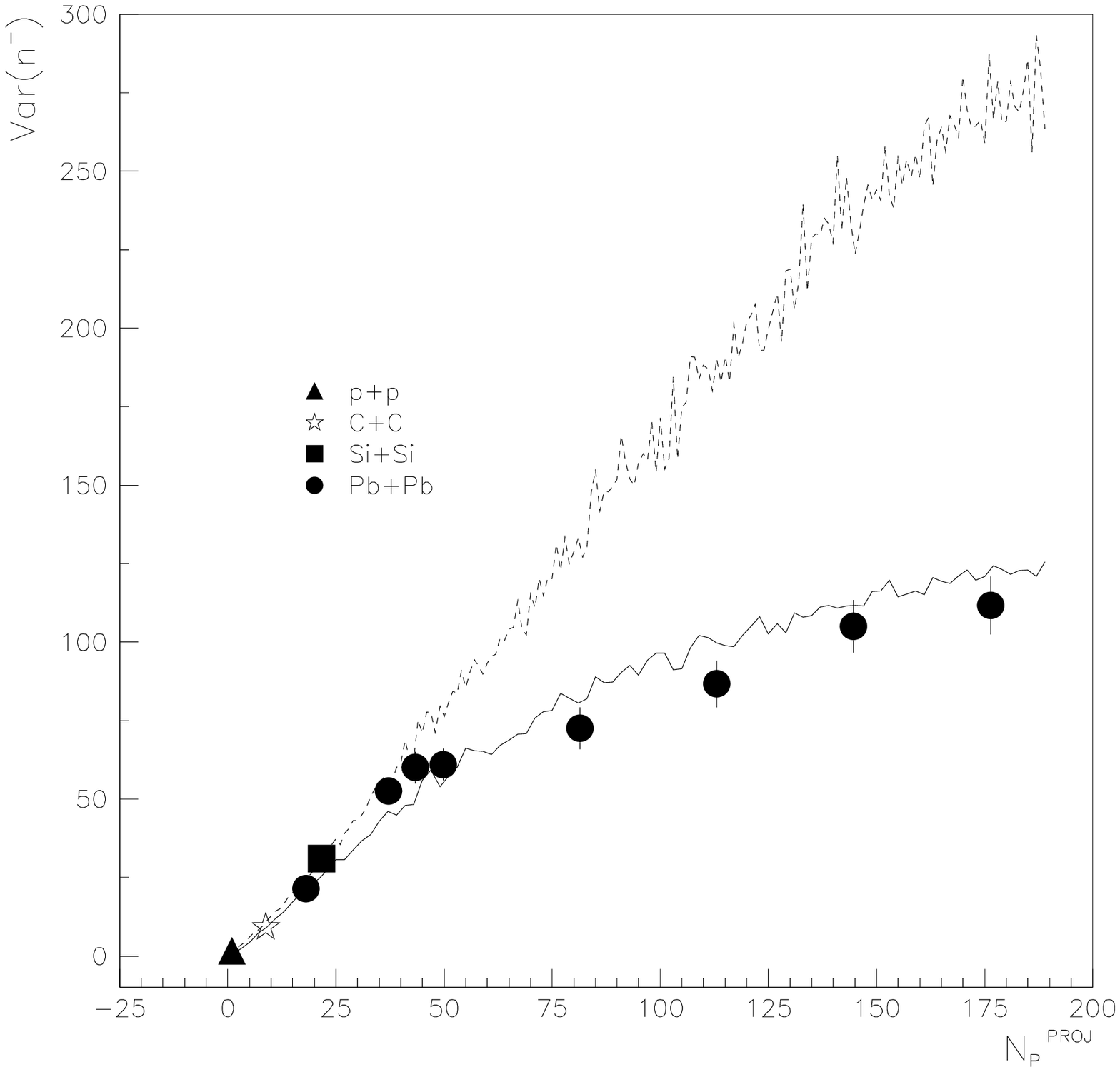}
\vskip 0.5cm
\caption{
Our results for the variance of negatively charged particles 
in Pb+Pb collisions at $P_{lab}=$158 AGeV/c
compared to NA49
experimental data. The dashed line corresponds to our result when clustering formation
is not included, the continuous line takes into account clustering.}
\label{fig2}
\end{figure}

\begin{figure}
\centering\leavevmode
\epsfxsize=6in\epsfysize=6in\epsffile{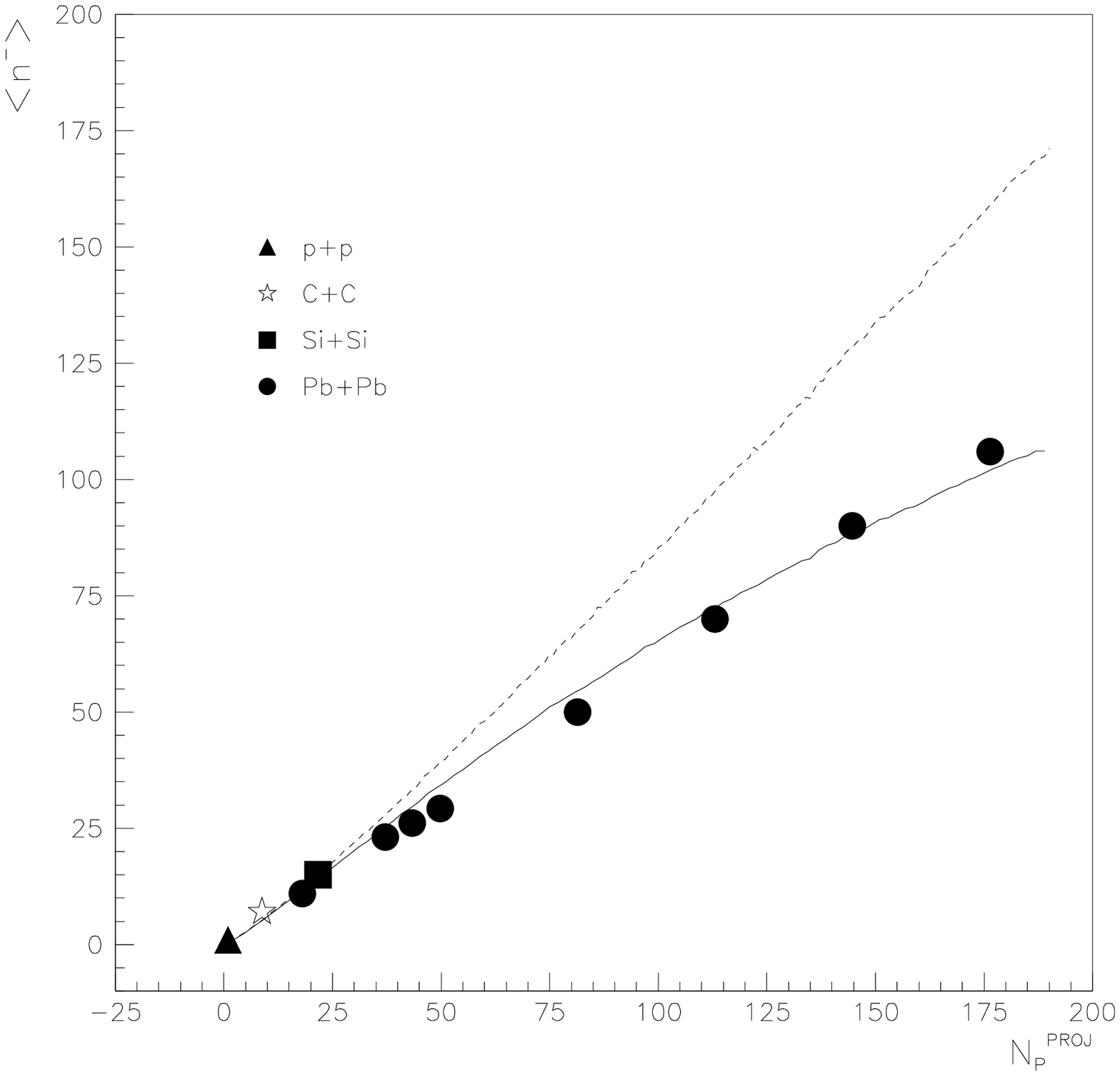}
\vskip 0.5cm
\caption{
Our results for the mean multiplicity of negatively charged particles 
in Pb+Pb collisions at $P_{lab}=$158 AGeV/c
compared to NA49
experimental data. The dashed line corresponds to our result when clustering formation
is not included, the continuous line takes into account clustering.}
\label{fig3}
\end{figure}

In Figs. \ref{fig2} and \ref{fig3} 
we present separately our results for the variance $V(n^-)$ and the mean multiplicity 
$<n^->$ of negatively charged particles. We have included our results without clustering formation. 
One can observe that, when clustering is included, we find a perfect agreement with experimental data 
for the mean multiplicity. Concerning the variance and the scaled variance, the agreement is less good, but
still one can see that the clustering works in the right direction: it produces a decrease of the variance 
in the central region --where the density of strings increases so the clustering has a bigger effect--.
Instead of that, without clustering, the scaled variance tends to a monotonic behaviour with centrality.
Note that, if no clustering is taken into account, our result for the variance is qualitatively similar to the
HIJING
simulation.
From eqs. (\ref{ec4}) to (\ref{ec8}) one can also deduce what will be the behaviour of the scaled variance
if both positively and negatively particles are taken into account: 
there will be an increase of the scaled variance in the fragmentation region --low number of participants and
low density of strings-- according to eq. (\ref{ec7}), due to the increase of $<\mu>_1$, that now becomes
proportional to $2/3$ of $<\mu>_0$. In the most central region our result for the scaled variance essentially
does not
change, since the dependence on $<\mu>_1$ 
is in this region much smaller, according to eq. (\ref{ec8}).
In our approach, the scaled variance for the positive particles is equal to the one for the negatives
particles, since both depend on $<\mu>_1$ in the same way.
This is in agreement with experimental data \cite{RybczynskiNA49}.

\begin{figure}
\centering\leavevmode
\epsfxsize=6in\epsfysize=6in\epsffile{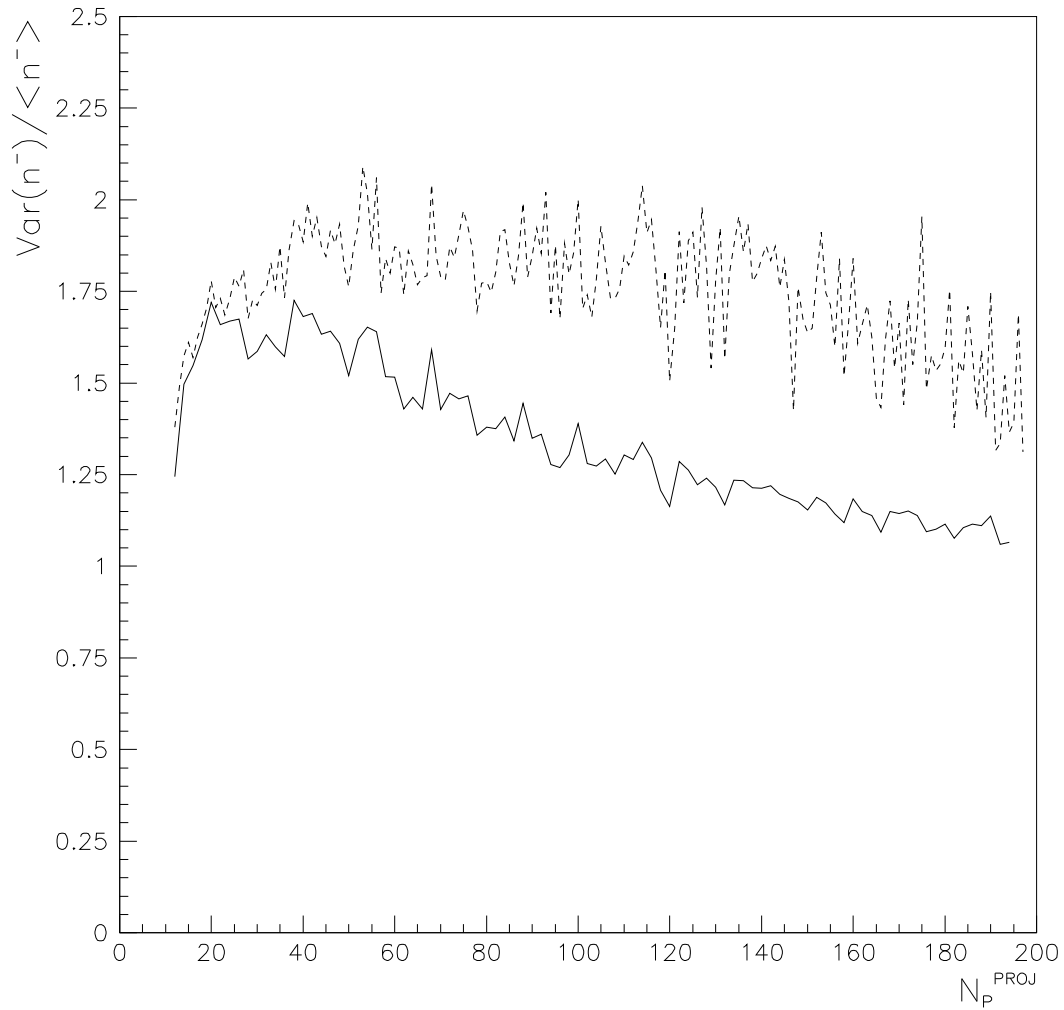}
\vskip 0.5cm
\caption{Our results for the scaled variance of negatively charged particles 
in Au+Au collisions at $\sqrt{s}=200$ GeV. The dashed line corresponds to our result when clustering formation
is not included, the continuous line takes into account clustering.}
\label{fig4}
\end{figure}

In Fig. \ref{fig4} we present our prediction for the scaled variance at RHIC energies. 
The behaviour is similar to
the one obtained at SPS energies. This is in accordance with our results for the mean $p_T$ fluctuations.
Note that now $<\mu>_1$ is going to be smaller that in the SPS case, since we take $\Delta y$=0.7, according
with the experimental acceptance of PHENIX experiment. This in principal implicates smaller correlations.
On the other hand, at RHIC energies we have a higher
value for the mean number of strings at fixed $N_{part}$. Both effects tend to compensate each other,
specially in the
small and mid centrality region --where $<\mu>_1$ plays a fundamental role, according to eq. (\ref{ec7})--.
In the large centrality region we can observe that the effect of clustering 
leads to a scaled variance very close to one.

In conclusion, we have found a non-monotonic dependence of the multiplicity fluctuations 
with the number of participants. The centrality behaviour of these fluctuations is very similar to the one 
previously found for the mean $p_T$ fluctuations.
In our approach, the physical mechanism responsible for multiplicity and mean $p_T$ fluctuations 
is the same \cite{Ferreiro:2003dw}:
the formation of clusters of strings
that introduces correlations between the produced particles. 
On the other hand, the mean $p_T$ fluctuations have been also attributed 
\cite{Adler:2003xq} to jet production in peripheral events, combined with jet suppression in more central
events. However, this hard-scattering 
interpretation, based on jet production and jet suppression, 
can not be applied to SPS energies, so it does not explain
the non-monotonic behaviour of the mean $p_T$ fluctuations 
neither the relation between mean $p_T$ and multiplicity fluctuations at SPS energy.
Other possible mechanism, extensively discussed
in \cite{Mrowczynski2004,Rybczy2004:9} are: combination of strong and electromagnetic interaction,
dipole-dipole interaction and non-extensive thermodynamics. 
Still, it is not clear if these fluctuations have
a kinematic or dynamic origin, but clustering of colour sources remains a good possibilty, since:

\begin{itemize}

\item
It can reproduce the qualitative behaviour of the even-by-event fluctuations with centrality.

\item 
In this approach, mean $p_T$ fluctuations and multiplicity fluctuations are naturally related.

\item It applies at SPS and RHIC energies.

\end{itemize}

\vskip 0.5cm

\noindent {Acknowledgments} \par\nobreak
It is a pleasure to thank N. Armesto for interesting discussions and
helpful suggestions.

\end{document}